\documentclass{amsart}
\usepackage{amsfonts,amsmath,amssymb}
\usepackage{multirow}
\usepackage[longnamesfirst,round]{natbib}
\usepackage{graphicx}
\usepackage{lineno,hyperref}
\modulolinenumbers[5]
\usepackage{mathtools}
\usepackage{diagbox}
\usepackage{setspace}

\newcommand{\e}{\mathbb{E}}
\newcommand{\var}{\mathbb{V}}

\newcommand{\bs}{\boldsymbol}
\newcommand{\mbf}{\mathbf}

\newcommand{\nrm}{\mathcal{N}}

\newcommand{\unif}{\mathcal{U}}

\newcommand{\poi}{\mathcal{P}}

\newcommand{\bY}{\bs{Y}}

\newcommand{\bZ}{\bs{Z}}
\newcommand{\bU}{\bs{U}}

\newcommand{\bu}{\bs{u}}

\newcommand{\bW}{\bs{W}}
\newcommand{\bzero}{\bs{0}}

\newcommand{\bomega}{\bs{\omega}}

\newcommand{\bzeta}{\bs{\zeta}}
\newcommand{\bpsi}{\bs{\psi}}

\newcommand{\btheta}{\bs{\theta}}

\newcommand{\bmu}{\bs{\mu}}

\newcommand{\momega}{\mbf{\Omega}}

\newcommand{\id}{\mbf{I}}
\newcommand{\reals}{\mathbb{R}}
\newcommand{\meat}{\bs{\mathcal{J}}}

\newcommand{\vscore}{\bs{\mathcal{V}}}

\newcommand{\ce}{\textsc{ce}}

\newcommand{\dt}{\textsc{dt}}

\newcommand{\bull}{\text{\raisebox{1pt}{\scalebox{.6}{$\bullet$}}}}

\title{On the Occasional Exactness of the Distributional Transform Approximation for Direct Gaussian Copula Models with Discrete Margins} 

\author{John Hughes\\
Department of Statistics\\The Pennsylvania State University}
\email{drjphughesjr@gmail.com}

\begin{document}

\begin{abstract}
The direct Gaussian copula model with discrete marginal distributions is an appealing data-analytic tool but poses difficult computational challenges due to its intractable likelihood. A number of approximations/surrogates for the likelihood have been proposed, including the continuous extension-based approximation (CE) and the distributional transform-based approximation (DT). The continuous extension approach is exact up to Monte Carlo error but does not scale well computationally. The distributional transform approach permits efficient computation but offers no theoretical guarantee that it is exact. In practice, though, the distributional transform-based approximate likelihood is so very nearly exact for some variants of the model as to permit genuine maximum likelihood or Bayesian inference. We demonstrate the exactness of the distributional transform-based objective function for two interesting variants of the model, and propose a quantity that can be used to assess exactness for experimentally observed datasets. Said diagnostic will permit practitioners to determine whether genuine Bayesian inference or ordinary maximum likelihood inference using the DT-based likelihood is possible for a given dataset.
\end{abstract}


\maketitle

\section{Introduction} 
\label{intro}

This article concerns maximum likelihood and Bayesian inference for direct Gaussian copula models for discrete outcomes. By `direct' we mean that the copula is applied to the response vector, as opposed to being applied at the second stage of the model (to the mean vector, for example). The stochastic form of the direct Gaussian copula model is given by
\begin{align}
\label{direct}
\nonumber\bZ = (Z_1,\dots,Z_n)'  & \; \sim\;  \nrm\{\bzero,\momega(\bomega)\}\\
\nonumber U_i = \Phi(Z_i) & \;\sim\; \unif(0,1)\;\;\;\;\;\;\;\;\;\;(i=1,\dots,n)\\
Y_i = F_i^{-1}(U_i) & \;\sim\; F_i(y\mid\bpsi),
\end{align}
where $\nrm$ denotes a Gaussian distribution, $\momega(\bomega)$ is a correlation matrix whose entries are functions of $\bomega\in\reals^q$, $\Phi$ is the standard Gaussian cdf, $\unif$ denotes a continuous uniform distribution, and $F_i(y\mid\bpsi)$, having parameters $\bpsi\in\reals^p$, is the cdf for the $i$th outcome $Y_i$. Note that $\bU=(U_1,\dots, U_n)'$ is a realization of the Gaussian copula $C_{\momega}(\bu)=\Phi_{\momega}\{\Phi^{-1}(u_1),\dots,\Phi^{-1}(u_n)\}$, which is to say that the $U_i$ are marginally standard uniform and exhibit the Gaussian correlation structure defined by $\momega$. Since $U_i$ is standard uniform, applying the inverse probability integral transform to $U_i$ in the final stage produces outcome $Y_i$ having the desired marginal distribution $F_i$.

We contrast the direct model with the hierarchical Gaussian copula model, which uses the copula to induce dependence among the outcomes by inducing dependence in, for example, the mean vector of the response. The hierarchical model is given by
\begin{align}
\label{hierarchical}
\nonumber\bZ = (Z_1,\dots,Z_n)'  & \; \sim\;  \nrm\{\bzero,\momega(\bomega)\}\\
\nonumber U_i = \Phi(Z_i) & \;\sim\; \unif(0,1)\;\;\;\;\;\;\;\;\;\;(i=1,\dots,n)\\
\nonumber \mu_i = F_i^{-1}(U_i) & \;\sim\; F_i(\mu\mid\bzeta),\\
Y_i  & \;\sim\; G_i(y\mid\mu_i,\bpsi),
\end{align}
where $\bzeta$ are marginal parameters for the mean vector, and cdf $G_i$ has mean parameter $\mu_i$ and other parameters $\bpsi$. In this scheme the mean vector $\bmu=(\mu_1,\dots,\mu_n)'$ carries the dependence structure of $\momega$, and the outcomes $Y_i$, which have marginal distributions $G_i$, are dependent because the marginal parameters $\mu_i$ are dependent. For example, a familiar hierarchical formulation for Poisson outcomes is the Gaussian-copula version of the gamma--Poisson model. For this model $\bmu$ would be a gamma random field, and $Y_i\sim\poi(\mu_i)\;(i=1,\dots,n)$, where $\poi$ denotes a Poisson distribution.

Although the hierarchical formulation enjoys certain advantages from a modeling point of view \citep{musgrove2016hierarchical} and will be more familiar to most readers, especially Bayesians, we favor the direct model because \citet{handeoliveira} found that, for point-level spatial data, the direct model is more flexible in terms of the range of feasible dependence, sensitivity to the mean structure, and modeling of isotropy. It stands to reason that the hierarchical model suffers from the same, or similar, limitations in other domains of application.

The above mentioned flexibility of the direct model comes at a price, however: for discrete outcomes, the likelihood is intractable. This has led to the development of a number of approximations/surrogates. In the rest of this article we will focus on two likelihood approximations, namely, the continuous extension (CE) and the distributional transform (DT). It is well known that the continuous extension, which is a Monte Carlo method, is exact up to Monte Carlo error and can be made practically exact by using a large Monte Carlo sample size (which is of course computationally burdensome). The distributional transform, by contrast, is computationally efficient but appears to be crude and is, in any case, never exact in theory. What is surprising about the distributional transform is that it is occasionally exact in practice. That is, for some sample sizes and interesting choices of $\momega$ and $\{F_i\}$, the DT-based objective function is so nearly equal to the true likelihood that said objective function can be used to do genuine maximum likelihood or Bayesian inference. We show this in the sequel.

The rest of this article is organized as follows. In Section~\ref{continuous} we present the likelihood for direct Gaussian copula models with continuous margins since the CE and DT objective functions are reminiscent of the likelihood for continuous outcomes. In Section~\ref{discrete} we present the likelihood for discrete outcomes, and explain why said likelihood is computationally intractable. In Sections~\ref{ce} and \ref{dt} we describe the CE and DT approximations, respectively, to the true likelihood for discrete marginals. In Section~\ref{simstudy} we verify by simulation that the DT approximation is effectively exact for some special cases of the model. In Section~\ref{realdata} we provide a means of discerning model misspecification for a given dataset. We conclude in Section~\ref{conclusion}.

\section{The likelihood for direct Gaussian copula models with continuous margins}
\label{continuous}

For correlation matrix $\momega(\bomega)$, continuous marginal cdfs $F_i(y\mid\bpsi)$, and marginal pdfs $f_i(y\mid\bpsi)$, the log-likelihood (corrrsponding to (\ref{direct}) above) of the parameters $\btheta=(\bomega',\bpsi')'$ given observations $\bY=(Y_1,\dots,Y_n)'$ is
\begin{align}
\label{loglik}
\ell_\textsc{ml}(\btheta\mid\bY)=-\frac{1}{2}\log\vert\momega\vert-\frac{1}{2}\bZ'(\momega^{-1}-\id)\bZ+\sum_i\log f_i(Y_i),
\end{align}
where $Z_i=\Phi^{-1}\{F_i(Y_i)\}$ and $\id$ denotes the $n\times n$ identity matrix. This objective function, being meta-Gaussian, presents no special computing challenges: the crux of obtaining the maximum likelihood estimate is the repeated evaluation of $\vert\momega\vert$ and $\momega^{-1}$---familiar challenges for anyone who has worked with elliptical distributions. We display (\ref{loglik}) only because the DT and CE approximations take forms that resemble (\ref{loglik}).

\section{The likelihood for discrete outcomes}
\label{discrete}

When the marginal distributions are discrete, the likelihood is given by
\begin{align}
\label{discretelike}
L(\bs{\theta}\mid\bY) &= \sum_{j_1=0}^1\dots\sum_{j_n=0}^1(-1)^kC_{\momega}(U_{1j_1},\dots,U_{nj_n}),
\end{align}
where $k=\sum_{i=1}^nj_i$, $U_{i0}=F_i(Y_i)$, and $U_{i1}=\lim_{y\nearrow Y_i}F_i(y)=F_i(Y_i^-)=F_i(Y_i-1)$. (Note that the last equality holds when the marginals have integer support, as they do in the remainder of this article.)

Unless $n$ is quite small, computation of (\ref{discretelike}) is infeasible because the multinormal cdf is unstable in high dimensions and because the sum contains $2^n$ terms. Thus a number of approximations/surrogates for (\ref{discretelike}) have been proposed. In this article we focus our attention on two approximations, one of which is based on the continuous extension \citep{Denuit:2005p951}, and the other of which is based on the distributional transform \citep{Ruschendorf:2009p1281}.

\section{The continuous extension}
\label{ce}

The continuous extension approach to maximum likelihood inference for Gaussian copula models with discrete marginals was developed by \citet{Madsen:2009p949}. The approach gets its name from a technique whereby a discrete random variable is transformed to a continuous one by introducing an auxiliary random variable supported on the unit interval \citep{Denuit:2005p951}.

To see how this can be accomplished, first suppose that $Y\sim F$ is a discrete random variable, and let $f$ be the pmf corresponding to $F$. Let $W$ be a continuous random variable supported on the unit interval, and suppose that $W$ has distribution function $G$, density function $g$, and is independent of $Y$. Then the continuation of $Y$ is the continuous random variable $Y^*=Y+(W-1)$. \citet{Denuit:2005p951} showed that $Y^*$ has distribution function $F^*(y)=F([y])+G(y-[y])f([y+1])$ and pdf $f^*(y)=g(y-[y])f([y+1])$, where $[\bull]$ returns the integer part of its argument. If we take $W$ to be standard uniform, $Y^*=Y-W$ and the distribution and density functions simplify to $F^*(y)=F([y])+(y-[y])f([y+1])$ and $f^*(y)=f([y+1])$, respectively.

For a direct Gaussian copula model with discrete margins we continue $\bY=(Y_1,\dots,Y_n)^\prime$ using $n$ independent standard uniforms $\bW=(W_1,\dots,W_n)^\prime$ and form the expected likelihood
\begin{align*}
L(\btheta\mid \bY) &\propto \e_{\bW}\left[\vert\momega\vert^{-1/2}\exp\left\{-\frac{1}{2}\bZ^{*\prime}(\momega^{-1}-\id)\bZ^*\right\}\prod_{i=1}^nf_i(Y_i)\right],
\end{align*}
where $\bZ^*=(\Phi^{-1}\{F_1^*(Y_1^*)\},\dots,\Phi^{-1}\{F_n^*(Y_n^*)\})^\prime$. Using a result proved by \citeauthor{Madsen:2010p1266}, one can show that this expectation is equal to the true likelihood given in (\ref{discretelike}).

We estimate the expectation using a sample-based approach. Let $m$ be a positive integer, and simulate a vector of independent standard uniforms, which \citeauthor{Madsen:2009p949} calls `jitters', $\bW_j=(W_{j,1},\dots,W_{j,n})^\prime$ for $j=1,2,\dots,m$. Then use the jitters to estimate the expected likelihood as
\begin{align}
\label{celike}
L_\ce(\btheta\mid \bY) &= \frac{1}{m}\sum_{j=1}^m\vert\momega\vert^{-1/2}\exp\left\{-\frac{1}{2}\bZ_j^{*\prime}(\momega^{-1}-\id)\bZ_j^*\right\}\prod_{i=1}^nf_i(Y_i),
\end{align}
where $\bZ_{j,i}^*=\Phi^{-1}\{F_i^*(Y_i-W_{j,i})\}$. This estimated likelihood can then be optimized to arrive at an approximate maximum likelihood estimate $\hat{\btheta}_\ce$ of $\btheta$.

Although the CE-based approach has the advantage of being exact up to Monte Carlo error, evaluation of (\ref{celike}) is computationally burdensome since a large number of jitters (at least 1,000, say) is typically required---so burdensome, in fact, that using the CE approach becomes infeasible as the sample size increases.

\section{The distributional transform}
\label{dt}

The distributional transform-based approximation was first proposed by \citet{Kazianka:2010p941} for fitting Gaussian copula geostatistical models.

It is well known that if $Y\sim F$ is continuous, $F(Y)$ has a standard uniform distribution. But if $Y$ is discrete, $F(Y)$ tends to be stochastically larger, and $F(Y^-)$ tends to be stochastically smaller, than a standard uniform random variable. This can be remedied by stochastically ``smoothing" $F$ at its jumps, a technique that goes at least as far back as \citet{Ferguson:1969p1279}, who used it in connection with hypothesis tests. More recently, the DT has been applied to stochastic ordering \citep{ruschendorf1981stochastically}, conditional value at risk \citep{burgert2006optimal}, and the extension of limit theorems for the empirical copula process to general distributions \citep{Ruschendorf:2009p1281}, for example.

Let $W\sim\mathcal{U}(0,1)$, and suppose that $Y\sim F$ and is independent of $W$. Then the distributional transform $G(W,Y)=(1-W)F(Y^-)+WF(Y)$ follows a standard uniform distribution and $F^{-1}\{G(W,Y)\}$ follows the same distribution as $Y$. See \citet{Ruschendorf:2009p1281} for a proof.

Turning back to the problem at hand, the DT-based approximate likelihood for direct Gaussian copula models with discrete marginals can be developed as follows. For each $i\in\{1,\dots,n\}$, let
\[
G_i(W_i,Y_i)=(1-W_i)F_i(Y_i^-)+W_iF_i(Y_i),
\]
where the $W_i$ are standard uniform random variables and are independent of one another and of the $Y_i$. Now put
\begin{align}
\label{ui}
 U_i&=\e_W\{G_i(W_i,Y_i)\mid Y_i\}=\{F_i(Y_i^-)+F_i(Y_i)\}/2=(U_{i0}+U_{i1})/2.
\end{align}
Then the approximate likelihood for our model is
\begin{align*}
L_\dt(\bs{\theta}\mid\bY) &= c_{\momega}(U_1,\dots,U_n)\prod_{i=1}^nf_i(Y_i),
\end{align*}
where $c_{\momega}$ denotes the copula density function. This implies the approximate log likelihood
\begin{align}
\label{likedt}
 \ell_\dt(\bs{\theta}\mid\bY) &= -\frac{1}{2}\log\vert\momega\vert-\frac{1}{2}\bZ^\prime(\momega^{-1}-\id)\bZ+\sum_{i=1}^n\log f_i(Y_i),
\end{align}
where $Z_i=\Phi^{-1}(U_i)$. Optimization of (\ref{likedt}) yields $\hat{\bs{\theta}}_\dt$.

Although the DT-based approximation appears to be almost ridiculously crude, the approximation performs well in a wide variety of circumstances and is even practically exact for some variants of the model (as we will demonstrate in the next section). Moreover, the DT approach does not entail the heavy computational burden of the CE approach.

\section{Verification by simulation}
\label{simstudy}

In this section we verify by simulation that $L_\dt$ is equivalent to the true likelihood for two realistic variants of the direct model. Since the true likelihood is unavailable, we use $L_\ce$ (with a large number of jitters) in place of the true likelihood, and compare the characteristics of $L_\dt$ to those of $L_\ce$.

\subsection{AR(1) process with negative binomial marginals}

One model for which $L_\dt=L$ is a Gaussian-copula version of an AR(1) process with negative binomial marginals. Specifically, let $1,\dots,200$ be the time indices at which we observe $Y_1,\dots,Y_{200}$, where $Y_i\;(i=1,\dots,200)$ is negative binomial with mean $\mu=12$. Let the dispersion parameter $k$ equal 7 so that $\var Y_i=\mu+\mu^2/k\approx 32.6$, where $\var$ denotes variance. For the AR(1) dependence structure we need $\momega_{ij}=\rho^{\vert i-j\vert}$ for time indices $i$ and $j$. For our simulation experiment we took $\rho=0.6$.

We simulated 1,000 datasets from this model. For each simulated dataset we optimized $\ell_\dt$ and $\ell_\ce$ to obtain $\hat{\btheta}_\dt$ and $\hat{\btheta}_\ce$, respectively, where $\btheta=(\rho,\mu,k)'$. (Note that we used 1,000 jitters for the CE procedure.) Then we computed the likelihood ratios $\Lambda_\dt(\btheta_0)=2\{\ell_\dt(\hat{\btheta}_\dt\mid\bY)-\ell_\dt(\btheta_0\mid\bY)\}$ and $\Lambda_\ce(\btheta_0)=2\{\ell_\ce(\hat{\btheta}_\ce\mid\bY)-\ell_\ce(\btheta_0\mid\bY)\}$, where $\btheta_0$ denotes the true value of $\btheta$ and $\bY=(Y_1,\dots,Y_{200})'$ denotes the sample. 

If either objective function is exact, we should expect its likelihood ratios to be $\chi^2(3)$ distributed. We used a Kolmogorov--Smirnov test to test this hypothesis for each procedure. The p-value for the DT ratios was 0.64, and the p-value for the CE ratios was 0.69. And so we fail to reject the null in both cases, i.e., the data are consistent with the hypothesis that they are $\chi^2(3)$ distributed. To provide further confirmation we carried out maximum likelihood estimation for both the two-parameter gamma distribution and the central $\chi^2$ distribution. For the DT ratios the gamma fit yielded an AIC of 4,104.6, the $\chi^2$ fit an AIC of 4,103.2. For the CE ratios the AIC values were 4,092.8 and 4,091.6. Additionally, the maximum likelihood estimates of the $\chi^2$ parameter were 3.05 for the DT ratios and 3.03 for the CE ratios, and the Wald confidence intervals were narrow and covered 3 for both datasets. Thus we choose the $\chi^2$ model in both cases.

\begin{figure}[ht]
   \centering
   \includegraphics[scale=.5]{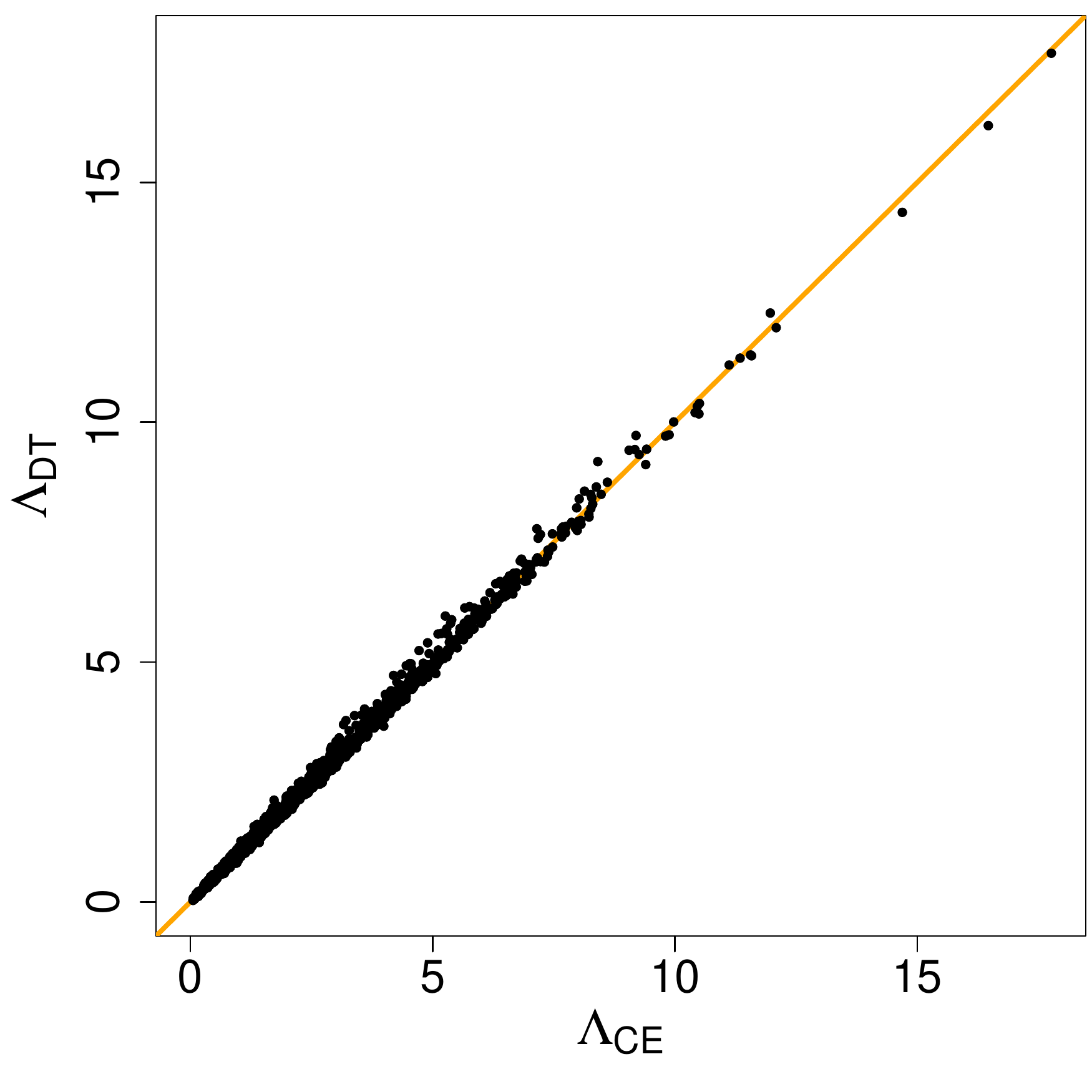}
   \caption{A plot of DT likelihood ratios versus CE likelihood ratios for 1,000 datasets simulated from an AR(1) model with negative binomial marginals. The line $y=x$ is shown in orange.}
   \label{fig:ratiosar}
\end{figure}

Having concluded that both objective functions are essentially exact, we should expect the DT ratios to agree with the CE ratios. We tested this hypothesis by applying Krippendorff's $\alpha$ to the ratios. The result was $\hat{\alpha}=0.9982$ with a 95\% bootstrap confidence interval equal to (0.9980, 0.9985). This of course implies near perfect agreement. Visual confirmation is provided by the plot shown in Figure~\ref{fig:ratiosar}.

\subsection{One-way mixed-effects ANOVA model with Poisson margins}

Another model for which $L_\dt=L$ is a Gaussian-copula generalization of the one-way mixed-effects ANOVA model. The model is given by
\begin{align}
\label{mod}
\nonumber\bZ &\;\sim\; \nrm\{\bzero,\momega(\omega=0.7)\}\\
\nonumber U_{ij} &\;=\; \Phi(Z_{ij})\;\;\;\;\;\;\;\;\;\;\;\;\;\;\;\;\;\;\;(i=1,\dots,20;\; j=1,2,3)\\
Y_{ij} &\;=\; F^{-1}(U_{ij}\mid\lambda=3),
\end{align}
where $\momega$ is block diagonal with blocks
\[
\momega_i=
\begin{pmatrix}
1 & 0.7 & 0.7\\
0.7 & 1 & 0.7\\
0.7 & 0.7 & 1
\end{pmatrix},
\]
and $F^{-1}(\cdot\mid 3)$ is the quantile function for the Poisson distribution with rate $\lambda=3$. This model could arise quite naturally in an effort to assess inter-rater reliability for count data, where an intraclass correlation of $\omega=0.7$ might be taken as evidence for substantial agreement among three raters for 20 units of analysis.

We simulated 1,000 datasets from this model, and once again optimized the DT and CE objective functions for each dataset as well as computing the likelihood ratios $\Lambda_\dt(\btheta_0)$ and $\Lambda_\ce(\btheta_0)$, where $\btheta_0=(\omega=0.7,\lambda=3)'$. And, since the model's parameter is two dimensional, for the sake of visual comparison we computed $L_\dt$ and $L_\ce$ for a single dataset on a $100\times 100$ grid (note that high-precision floating-point arithmetic was required to avoid underflow \citep{rmpfr}). 

\begin{figure}[ht]
   \centering
   \begin{tabular}{c}
   \includegraphics[scale=.5]{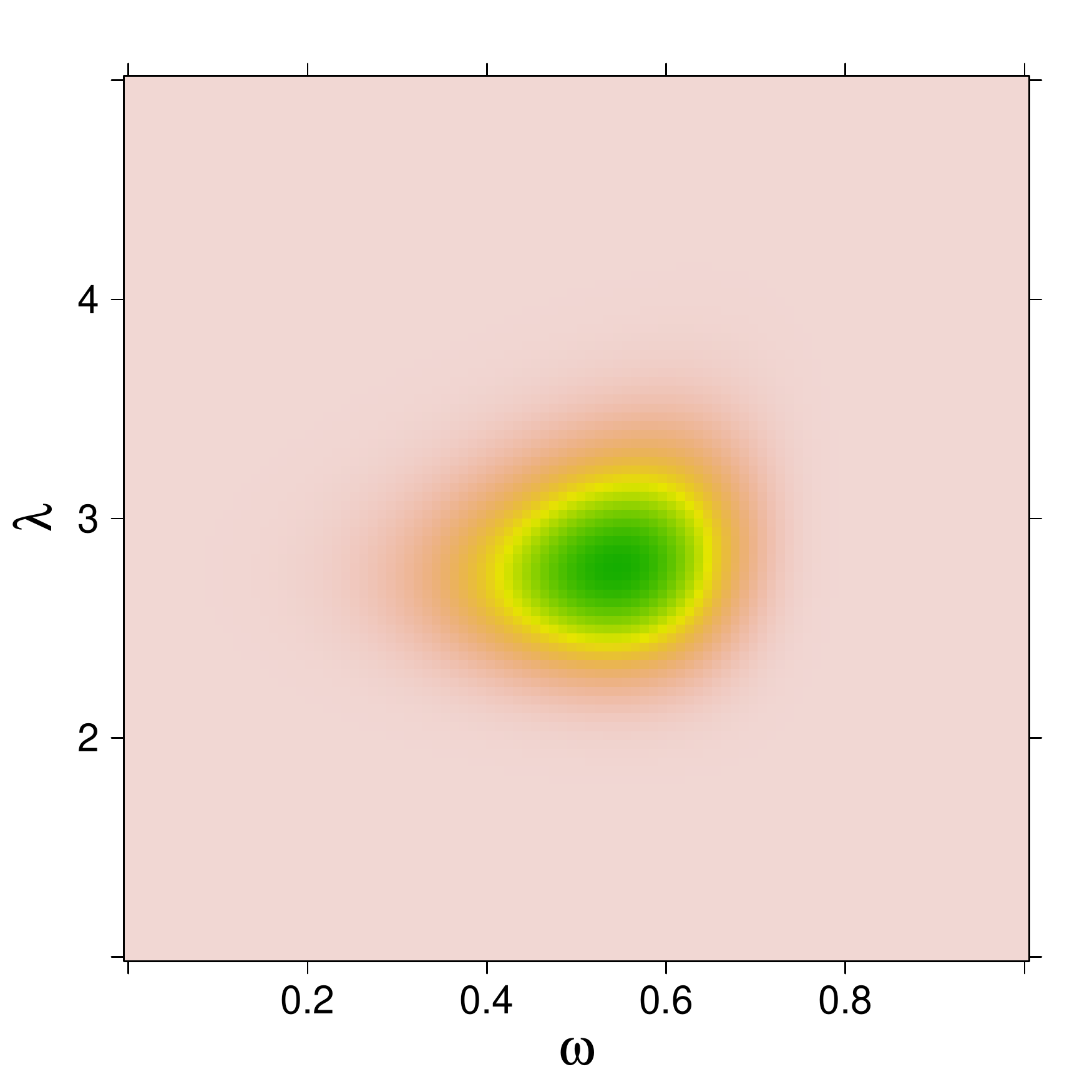}\\
   \includegraphics[scale=.5]{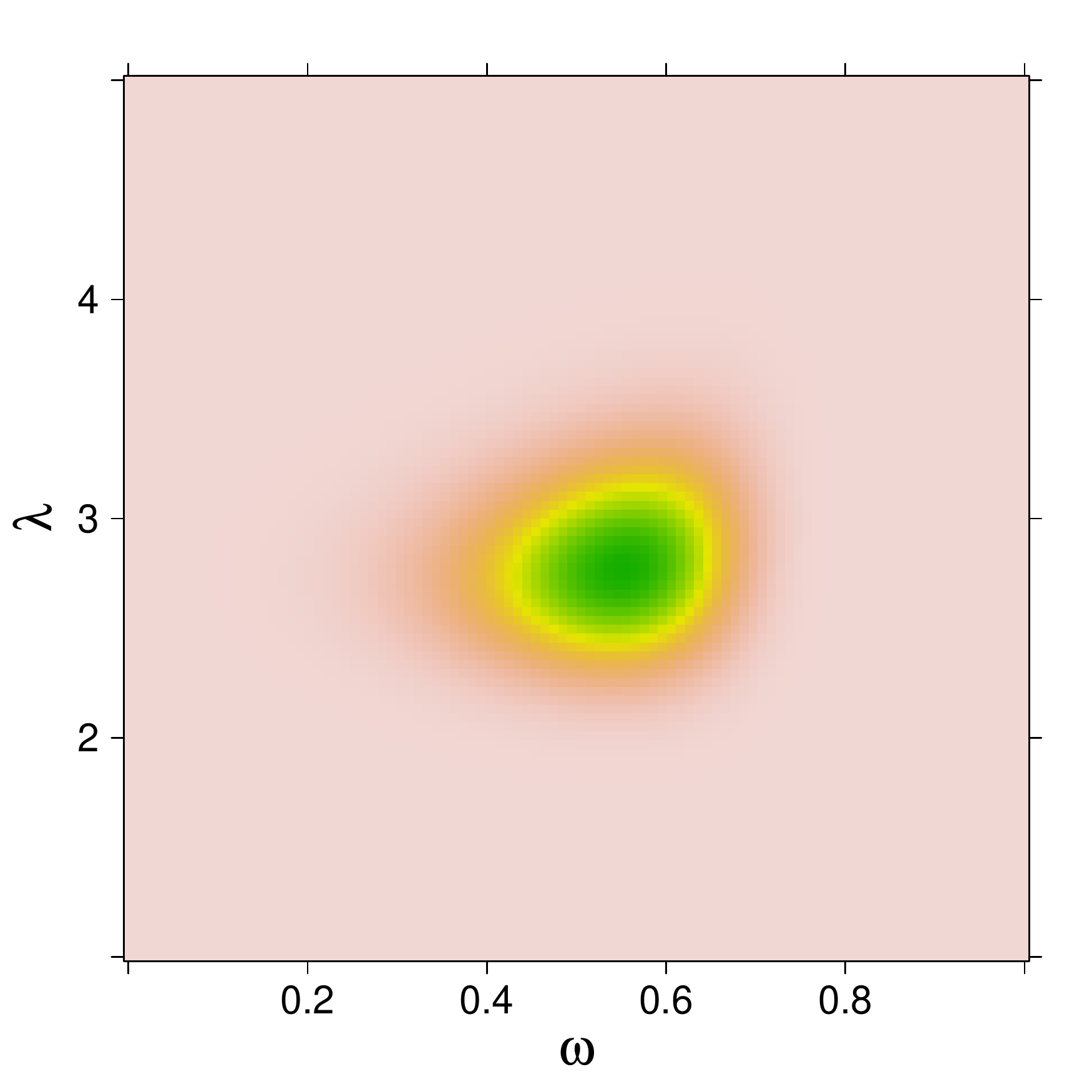}
   \end{tabular}
   \caption{$L_\ce$ (top) and $L_\dt$ for a single dataset simulated from the one-way mixed-effects ANOVA model with Poisson marginals.}
   \label{fig:like}
\end{figure}

\begin{figure}[ht]
   \centering
   \includegraphics[scale=.5]{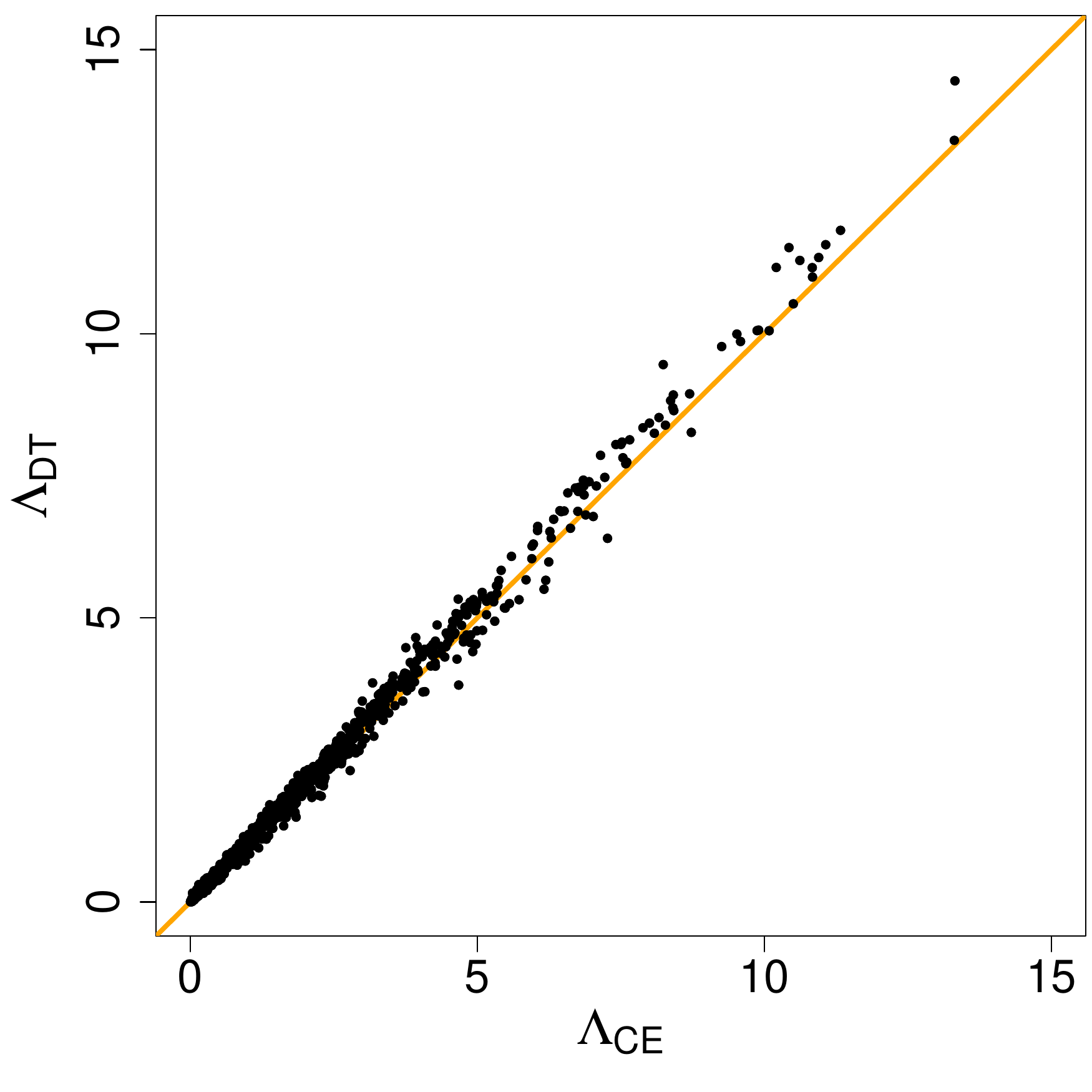}
   \caption{A plot of DT likelihood ratios versus CE likelihood ratios for 1,000 datasets simulated from a one-way mixed-effects ANOVA model with Poisson marginals. The line $y=x$ is shown in orange.}
   \label{fig:ratiosrater}
\end{figure}

Level plots of $L_\dt$ and $L_\ce$ are shown in Figure~\ref{fig:like}. We see that the two objective functions are practically indistinguishable. And the likelihood ratios (shown in Figure~\ref{fig:ratiosrater}) once again exhibit very high agreement (Krippendorff's $\hat{\alpha}=0.996$) and have the expected $\chi^2(2)$ distribution.

\section{A useful diagnostic quantity for experimentally observed data}
\label{realdata}

In this section we describe a quantity that can be used to assess the exactness of $L_\dt$ for a given dataset. This quantity should prove appealing to practitioners since the quantity is intuitive and can be computed efficiently.

From the theory of maximum likelihood we know that, in many scenarios, the second Bartlett identity,
\[
\meat(\btheta_0)=\vscore(\btheta_0),
\]
fails to hold if the model is misspecified \citep{bartlett1953,white}, where $\meat(\btheta_0)=-\e\nabla^2\ell_\dt(\btheta_0)$ is the negated expected curvature of the objective function, and $\vscore(\btheta_0)=\e\nabla\nabla'\ell_\dt(\btheta_0)$ is the variance of the score function. This implies that
\[
\kappa\vcentcolon=\Vert\meat-\vscore\Vert_\text{F}=0,
\]
where $\Vert\cdot\Vert_\text{F}$ is the Frobenius norm. Thus the quantity $\hat{\kappa}=\Vert\hat{\meat}-\hat{\vscore}\Vert_\text{F}$ is useful for assessing the exactness of $L_\dt$ for a given dataset: a value close to zero suggests that $L_\dt=L$ for the data in question.

We use a (parallel) parametric bootstrap to estimate $\vscore$:
\[
\hat{\vscore}(\hat{\btheta}_\dt)=\frac{1}{n_b}\sum_{i=1}^{n_b}\nabla\nabla'\ell_\dt(\hat{\btheta}_\dt\mid\bY^{(i)}),
\]
where $n_b$ is the bootstrap sample size and $\bY^{(i)}$ is the $i$th sample simulated from the model at $\btheta=\hat{\btheta}_\dt$. We take as our estimate of $\meat$ the Hessian that is produced as a side-effect of optimizing $\ell_\dt$, or we produce a bootstrap estimate of $\meat$ along with $\hat{\vscore}$:
\[
\hat{\meat}(\hat{\btheta}_\dt)=-\frac{1}{n_b}\sum_{i=1}^{n_b}\nabla^2\ell_\dt(\hat{\btheta}_\dt\mid\bY^{(i)}).
\]



Our $\kappa$ diagnostic could also be used to explore a region of the parameter space for a given model, perhaps revealing multiple parameter values for which $L_\dt$ is an adequate replacement for the true likelihood. We applied this technique in the context of the one-way mixed-effects ANOVA model with Poisson marginals. Specifically, for a two-way factorial design with $\lambda=1,2,3,4$ and $\omega=0.6,0.7,0.8,0.9$, we computed $\hat{\kappa}$ using 10,000 simulated datasets at each of the 16 design points. The resulting $\hat{\kappa}$ values are shown in Table~\ref{tab:kappa}.

\begin{table}[ht]
   \centering
   \begin{tabular}{c|cccc}
     \backslashbox{$\lambda$}{$\omega$}    &  0.6 & 0.7 & 0.8 & 0.9\\\hline
1 & 35 & 62 & 208 & 6,732\\
2 & 10 & 11 & 72 & 3,016\\
3 & 2 & 7 & 35 & 1,544\\
4 & 1 & 5 & 25 & 875
   \end{tabular}
   \caption{Using the $\kappa$ diagnostic to explore the parameter space of the one-way mixed-effects ANOVA model with Poisson marginals.}
   \label{tab:kappa}
\end{table}

We see an illuminating and predictable interaction between the marginal variance and the dependence strength. For a given value of $\lambda$, the quality of $L_\dt$ as a replacement for the true likelihood decreases as the dependence strength increases. And for a given value of $\omega$, increasing $\lambda$ (and hence the marginal variance) brings $L_\dt$ ever closer to the true likelihood. As expected, choosing $\btheta_0=(\omega=0.7,\lambda=3)'$ leads to a small value of the diagnostic quantity. By contrast, $L_\dt$ is clearly not a suitable substitute for the true likelihood when $\btheta_0=(\omega=0.9,\lambda=1)'$, for example.

\section{Conclusion}
\label{conclusion}

In this article we showed that the distributional transform-based objective function for direct Gaussian copula models with discrete margins is sometimes effectively exact, in which case true Bayesian inference is possible. We demonstrated said exactness for two interesting variants of the model: an AR(1) process with negative binomial marginals, and a one-way mixed-effects ANOVA model with Poisson marginals. Then we developed a diagnostic quantity based on Bartlett's second identity. This quantity, which can be used to assess the exactness of the DT-based objective function for experimentally observed datasets, is intuitive and can be computed in embarrassingly parallel fashion. This diagnostic procedure will allow practitioners to determine whether $L_\dt$ can be used to do genuine Bayesian inference or ordinary maximum likelihood inference for a given dataset.

\section*{Acknowledgement}

The author is grateful to Ben Seiyon Lee for helpful discussions regarding this work.


\bibliography{refs}
\bibliographystyle{apalike}

\end{document}